# Photostriction Facilitates Relaxation of Lattice Distortion in Two-Dimensional Perovskites

Jin Zhang[1,*], Kun Yang[1], Jianxin Yu[1], Jia Zhang[2], Sheng Meng[3,4], Xinghua Shi[1], Wei-Hai Fang[5]

[1]Laboratory of Theoretical and Computational Nanoscience, National Center for Nanoscience and Technology, Chinese Academy of Sciences. Beijing 100190, China

[2]Max Born Institut für Nichtlineare Optik und Kurzzeitspektroskopie, Berlin 12489, Germany

[3]Beijing National Laboratory for Condensed Matter Physics, and Institute of Physics, Chinese Academy of Sciences, Beijing 100190, P. R. China

[4]Songshan Lake Materials Laboratory, Dongguan, Guangdong 523808, China

[5]College of Chemistry, Key Laboratory of Theoretica l & Computational Photochemistry of Ministry of Education, Beijing Normal University, Beijing 100875, China.

* Corresponding author: Jin Zhang (jinzhang@nanoctr.cn)

# Abstract

The photostriction effect, a light-induced mechanical deformation in materials, originates from the intricate interplay between lattice structure and electronic excitation. In photovoltaic semiconductors, this effect plays a crucial role in shaping non-equilibrium structural responses, yet its fundamental mechanism remains elusive. Here, we uncover lattice expansion and structural reconfiguration in two-dimensional (2D) perovskites driven by photoinduced excitation using first-principles calculations. Our findings reveal that the photoinduced carriers can lead to a substantial lattice expansion by about 2%. The expanded lattice facilitates strain relaxation with the amplitude of 20% by increasing interatomic distances and reducing internal stresses, thereby enhancing structural stability. The lattice dynamics can be systematically engineered through photodoping density, unveiling a new pathway to modulate light-matter interactions in 2D perovskites. These insights not only advance the understanding of optically driven structural dynamics but also offer a guiding principle for optimizing next-generation high-efficiency photovoltaic devices and optoelectronics.

Hybrid organic-inorganic perovskites, with remarkable efficiency, tunable property, and cost-effective fabrication, have emerged as a family of promising photovoltaic materials for sustainable energy harvesting [1-11]. Recent advancements in material engineering and interfacial optimization have propelled perovskite solar cells to achieve power conversion efficiencies exceeding 26% [7-9]. To address challenges related to structural stability, two-dimensional (2D) perovskites have been developed by incorporating layered structures with alternating organic components [12-16]. The 2D perovskites offer improved environmental stability, enhanced moisture resistance, and versatile optoelectronic properties, facilitating their use as strong candidates for durable, long-lasting photovoltaic applications [17-32].

Light-induced structural dynamics in the hybrid perovskites unlock new opportunities for improving stability and performance, potentially enabling reversible modifications that directly influence their optoelectronic properties [2-4]. The photostriction effect describes the phenomenon where the lattice structures undergo significant changes under the condition of optical excitation, which is beneficial for flexible optoelectronic devices [33-34]. Recently, Zhang and collaborators observed an ultrafast relaxation of lattice strain in 2D perovskites, along with a light-induced reduction of lattice distortions, attributed to strong coupling between the electron-hole plasma and lattice structures [2]. In addition, Tsai *et al.* demonstrated that photoinduced lattice expansion improves perovskite solar cell efficiency from 18.5% to 20.5%, paving the way for uncovering novel structural phases and dynamic behaviors in hybrid perovskites [3]. It has also been unveiled that giant photostriction plays a role in organic-inorganic lead halide perovskites and implies for understanding the exceptional photovoltaic performance [4]. However, the interaction between photoexcited carriers and lattice structures, particularly the lattice distortion caused by the photostriction effect, remains unclear.

Here, we employ time-dependent density functional theory (TDDFT) to unveil the atomic mechanism governing photostriction and structural deformations in 2D perovskites. Our simulations reveal that photoinduced carriers drive a substantial lattice expansion and significant relaxation of lattice distortions, highlighting the intricate interplay between electronic excitations and lattice dynamics. By systematically tuning the photodoping density, a direct modulation of the structural response is observed, providing new insights into light-matter interactions. These findings establish a fundamental framework for understanding optically driven structural dynamics and provide guiding principles for optimizing the design of next-generation high-efficiency optoelectronic devices.

Two-dimensional perovskites consist of alternating organic and inorganic layers, offering improved stability against moisture and thermal degradation [27]. The materials exhibit direct bandgaps suitable for solar energy conversion and demonstrate high absorption coefficients in the visible spectrum, making them excellent candidates for light-harvesting applications. 2D perovskites can be categorized into Dion-Jacobson (DJ) and Ruddlesden-Popper (RP)phases [1-2, 13-16]. The DJ-type perovskites consist of diammonium cations that form stronger covalent or hydrogen bonds while the RP-type perovskites typically feature single-ammonium spacer layers with van der Waals gaps between layers. This enhanced connectivity in DJ-type structures facilitates better charge transport across layers, overcoming the carrier mobility limitations in RP-type perovskites. With the combination of stability, efficient charge transport, and strong optical absorption, 2D perovskites present a promising alternative to traditional perovskite solar cells, paving the way for durable and high-performance photovoltaic devices [12-16].

In the DJ-type perovskites, the interlayer spacer ligands are maintained by specific cations, *e.g.*, divalent 4-aminomethylpiperidinium (4AMP), which replace the conventional monovalent spacer cations found in RP-type perovskites [1-2]. The structure features corner-sharing octahedra, forming the 2D perovskite-like network. The incorporation of methylammonium (MA) within the inorganic slabs influences the lattice distortion and electronic properties, while the rigid cations enforce a well-defined interlayer spacing. Especially, (4AMP) MAPb$_2$I$_7$ with n=2 (here, n denotes the number of inorganic layers) is a representative DJ-phase perovskite that has gained significant interest for the solar cells.

As shown in Figure 1, the atomic structure of the 2D perovskite consists of the layered perovskite framework, where inorganic slabs are separated by organic cations (only Pb-I frameworks are presented for clarity). We introduce an order parameter 2θ as the interlayer Pb-I bond tilt to quantify the lattice distortion (Figure 1a). For the fully relaxed structural distortion of (4AMP) MAPb$_2$I$_7$, the value of lattice distortion is approximately 2θ=24°, whereas the undistorted phase exhibits no structural modulation, corresponding to 2θ=0° (Figure 1b). To illustrate the intermediate atomic configurations, the linear interpolation is utilized to generate a series of geometries between the two reference structures as 2θ=24° and 0°. The intermediate geometries are utilized to compute energy barriers at varying carrier doping levels.

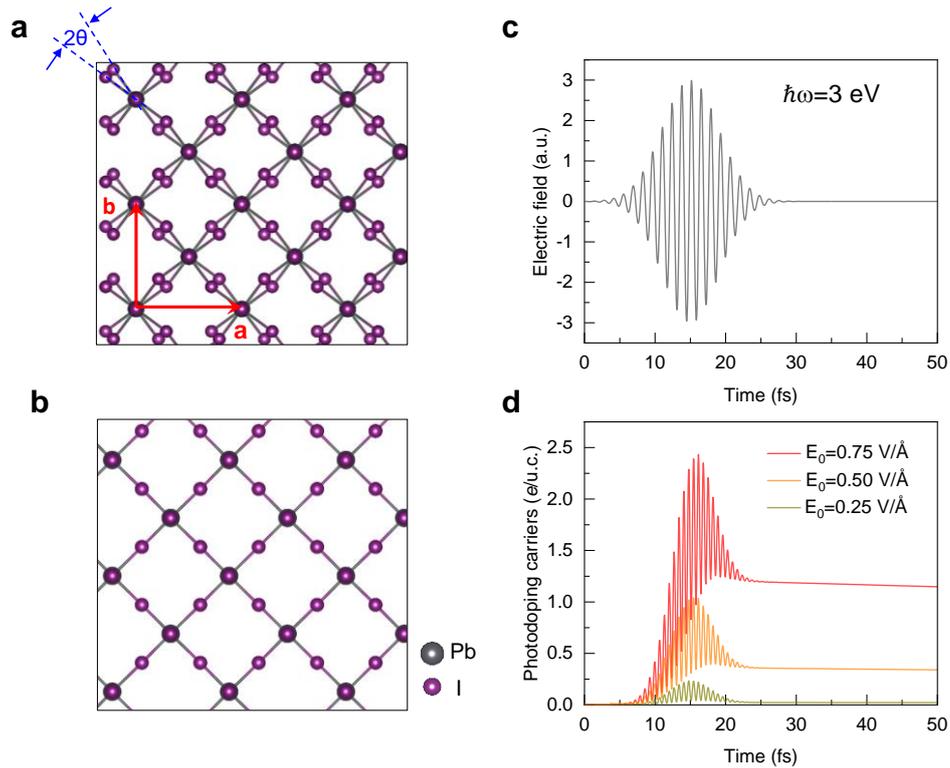

**Figure 1. Lattice structures and photoexcited carrier dynamics in 2D perovskites.** (a) Atomic structure of fully relaxed 2D perovskite (4AMP) MAPb$_2$I$_7$. The lattice distortion is illustrated with the angle (2θ), as illustrated in blue. Large black and small purple spheres denote Pb and I atoms, respectively. Only Pb-I frameworks are shown for clarity. The in-plane lattice constants are labeled as a and b in red. (b) Atomic structure of 2D perovskite without distortion for a reference structure (2θ=0°). The fully relaxed structure (2θ=24°) and the structure without distortions serve as the reference points, with intermediate geometries linearly interpolated. (c) Applied laser pulse with the photon energy of 3 eV. The shape of the laser field is shown in light grey. (d) The number of photodoping carriers excited from the valence to conduction bands for the laser strengths of E$_0$=0.25, 0.50 and 0.75 V/Å, respectively. The ionic structures are fixed during the dynamics.

We further perform real-time TDDFT simulations [35-40] to explore laser-driven phase transition in 2D perovskites, capturing the complex quantum interactions within the lattice and the excited carriers. As depicted in Figure 1c, the structural modifications induced by optical excitation are shown under different laser photon energies. To track the laser-excited dynamics in 2D perovskites, Gaussian-envelope functions are employed to characterize the laser pulses:

$$E(t) = E_0 \cos(\omega t) \exp\left[-\frac{(t-t_0)^2}{2\sigma^2}\right] \quad (1)$$

where $E_0$, $\omega$, $t_0$, and $\sigma$ are the maximum strength, the photon energy, the peak time of the electric field, and the width of the Gaussian pulse, respectively. Laser pulses with different strengths are employed to excite the system, and the ensuing electronic and structural responses are analyzed after photoexcitation. Under optical excitation, the electronic subsystem undergoes direct excitation, triggering structural dynamics via effective electron-phonon interactions.

**Optically induced photodoping effect.** To elucidate the laser-induced carriers, we thoroughly analyze the photoexcited carrier density in the system, specifically tracking the progression of excited electron-hole populations. As illustrated in Figure 1d, the number of excited electrons in the 2D perovskite follows the laser field obviously under varying strengths. At the end of the laser pulses (~50 fs), different fractions of electron-hole densities are excited from valence to conduction states. For $E_0$=0.25 V/Å, the total photodoping density reaches 0.02 e/u.c. after the laser pulse. This value increases to 0.36 e/u.c. at an intensity of $E_0$=0.50 V/Å. Notably, under a laser field of $E_0$=0.75 V/Å, the photodoping density further rises to 1.17 e/u.c., highlighting the nonlinear response of the system to increasing the field strength for the above-bandgap excitation with $\hbar\omega$=3 eV.

In addition, the real-time TDDFT calculations allow us to monitor the energetic distributions of the photoexcited carriers, which are available in the supplementary materials (Figure S1). The carrier occupations indicate that strong laser pulses proficiently excite a number of electrons and holes. Furthermore, the laser-induced carrier occupations are employed as the constraint on the energy analysis in orbital-constrained density functional theory calculations (See method for more details). The framework modifies the standard DFT method by incorporating additional constraints, effectively guiding the self-consistent field procedure towards the targeted electronic state. The approach is particularly useful to illustrate systems with fixed occupations or enforcing constraints such as charge or spin state configurations. This ensures that the system retains a predefined occupation configuration.

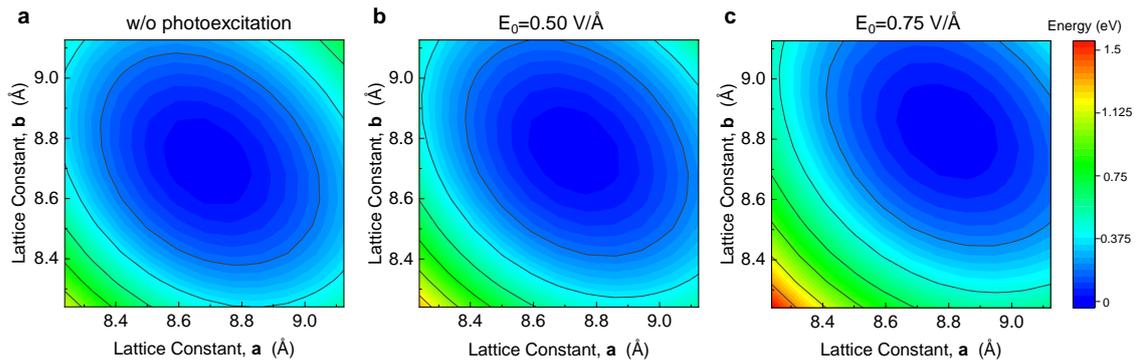

**Figure 2. Photodoping carrier induced lattice expansion in 2D perovskite.** (a) Energy diagram of the 2D perovskite (4AMP) MAPb$_2$I$_7$ with different lattice constants without photoexcitation. (b) The same quantity with (a) under the laser intensity of $E_0$=0.50 V/Å. (c) The same quantity with (a) under the laser intensity of $E_0$=0.75 V/Å. In all panels, **a** and **b** denote the lattice constants along the two directions.

**Photostriction effect in 2D perovskites.** Lattice expansion and structural distortion in the hybrid perovskites remarkably affect electronic properties, photoelectric conversion efficiency and stability [2-9]. In previous experiments, structural and device analyses reveal that light-induced lattice expansion improves the power conversion efficiency of a mixed-cation, pure-halide planar device [3].

We investigate photodoping-induced changes in lattice constants of the 2D perovskite. As shown in Figure 2a, the fully relaxed lattice constant of (4AMP) MAPb$_2$I$_7$ is 8.68 Å, in good agreement with the experimental observations [2]. Under laser excitation at $E_0$=0.50 V/Å, a slight expansion (~1%) is observed, increasing the lattice constant to 8.80 Å (Figure 2b). Further excitation at $E_0$=0.75 V/Å leads to a continued expansion to 8.90 Å, corresponding to a 2% increase (Figure 2c). The findings indicate that laser-induced carriers apparently influence lattice structures (*i.e.*, the photostriction effect), offering a pathway for further optimizing structural properties. The lattice expansion is capable of relaxing local lattice strain in the perovskites and thereby may improve device performance.

To elucidate the impact of lattice expansion on electronic modulation, we evaluated the band structures, revealing a clear correlation between lattice deformation and electronic properties, as shown in Figure S2. Under the condition of the lattice expansion of 1%, the bandgap increases to 1.685 eV. Furthermore, the value grows further to 1.699 eV for the amplitude of 2%. The observed bandgap increase highlights the crucial influence of lattice constants in modulating electronic properties, potentially impacting the device characteristics.

It is illustrated that light-driven lattice expansion improves perovskite cell performance significantly, opening new avenues for discovering unprecedented structural phases and dynamic properties in hybrid perovskites in experiments [3]. Moreover, Zhang and coauthors demonstrated that a mere 1% lattice expansion suppresses the nonradiative capture coefficient by an order of magnitude employing first-principles calculations [28]. This reduction is not driven by bandgap shifts or defect transition level changes but instead arises from enhanced defect relaxation associated with charge-state transitions in the expanded lattice. The light-induced lattice expansion is promising to improve perovskite device performance by alleviating local lattice strain and lowering energy barriers at perovskite-contact interfaces.

**Photostriction facilitates relaxation of structural distortion**. In order to investigate the impact of photodoping on structural distortions in 2D perovskites, we explore the characteristics of the energy landscape under different laser intensities. Figure 3 presents the calculated energy profiles of the system under different electric field strengths, revealing how the atomic structure evolves as a function of laser intensity. Specifically, under a laser intensity of $E_0$=0.25 V/Å (Figure 3a), the system reaches its minimum energy configuration at a diffraction angle of 2θ=23.3°, indicating a moderate level of structural distortion. As the laser intensity increases to $E_0$=0.50 V/Å (Figure 3b), the energy minimum shifts to 2θ=21.7°, suggesting a further relaxation in distortion. When the intensity is further elevated to $E_0$=0.75 V/Å, the minimum energy configuration locates at 2θ=20.0°, signifying an even greater relaxation of structural strain (Figure 3c). These results demonstrate a gradual decrease in structural distortion with increasing photodoping densities.

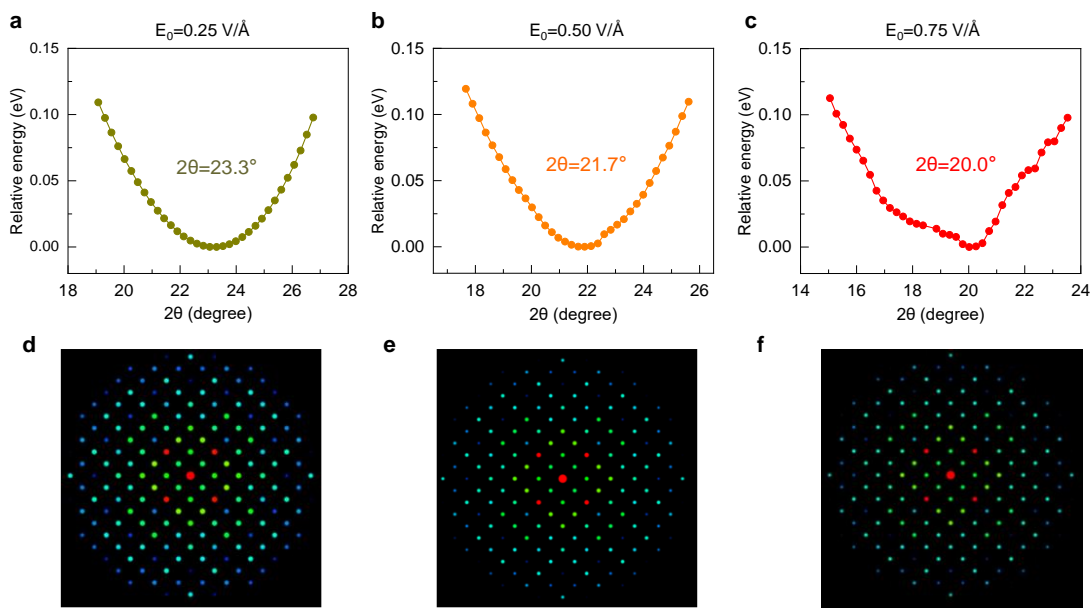

**Figure 3. Photodoping-induced relaxation of structural distortion.** (a) Energy profile of 2D perovskite under the laser intensity of $E_0$=0.25 V/Å. The energy minimum is located at 2θ=23.3°. (b) The same quantity for $E_0$=0.50 V/Å. The energy minimum is located at 2θ=21.7°. (c) Corresponding energy curve for the laser intensity of $E_0$=0.75 V/Å, showing a minimum at 2θ=20.0°. (d-f) Electron diffraction simulation for the corresponding structures. The linear interpolation is applied to obtain the transition path and atomic structures. The lattice constants are fixed at the fully relaxed values at the ground state. The indexing of Bragg planes is labeled as {hk0}. Diffraction peaks circled with red and green colors belong to the {220} and {400} Bragg peaks, respectively.

Furthermore, electron diffraction simulations are performed to analyze the corresponding structural evolution, as exhibited in Figures 3. Our results suggest that the diffraction pattern transformation is linked to light-induced in-plane lattice reorganization. Compared to the undisturbed lattice structure, the photodoping-induced structures exhibit a reduction in adjacent octahedral tilt along the stacking direction and improved alignment of iodine atoms in both the octahedral edge {h00} and diagonal {hh0} directions.

This alignment is particularly noticeable at the Pb-I bond spacing {220} and half the octahedral edge length {400} [2]. As shown in Figure 3d, the diffraction pattern at the ground state displays the high intensity at both {220} and {400} peaks, characterized by well-defined interference fringes, indicating a strong electron scattering signal. In Figure 3e-f, the intensity decreases noticeably by 50%, attributed to the considerable relaxation of lattice distortion. The observations thus support the finding that the light-induced relaxation of the lattice distortion in 2D perovskite arises from a collective lattice reorganization that minimizes octahedral rotations.

Additionally, the calculated band structures reveal a clear connection between the structure distortion and the electronic properties [Figure S3]. For the fully relaxed structure ($2\theta=24.0°$), the band structure features a direct bandgap of 1.649 eV on the PBE level. Under the photodoping-reduced condition ($2\theta=21.7°$), the bandgap decreases to 1.576 eV. Moreover, in the configuration with $2\theta=20.0°$, the bandgap further reduces to 1.480 eV. This trend suggests a gradual weakening of electronic confinement due to structural modulation, resulting in narrower bandgaps. The observed electronic structures underscore the critical role of lattice constants and structural distortions in tailoring the electronic properties, which have distinct implications for the optical and transport behaviors.

The real-time structural evolution using TDDFT method is subsequently analyzed under varying laser pulse excitations (Figure S4). Structural snapshots are recorded following above-bandgap excitation (3 eV), capturing the transient lattice configuration of the 2D perovskite system. For an electric field amplitude of $E_0=0.25$ V/Å, the structural distortion exhibits oscillatory behavior around 24°. In contrast, when photocarriers are not yet fully excited, the distortion initially exhibits a slight increase at the strengths of $E_0=0.5$ V/Å at early times (before 30 fs). Subsequently, the distortion angle undergoes a pronounced decrease in 60 fs, reaching the value of 23.1°. The results validate that the timescales of lattice distortion reduction are on several tens of femtoseconds from real-time TDDFT evolution, aligning well with the theoretical predictions as discussed above.

We note that extending the simulation to longer timescales presents significant computational challenges due to the large system size. The complexity arises from the need to integrate different degrees of freedom over extended timescale while maintaining numerical stability and accuracy. As the system size increases, the computational cost grows obviously, making direct and longer simulations infeasible. A comprehensive analysis requires not only prolonged simulations but also multiple trajectories to account for statistical variations and ensure convergence of the results. The analysis of the structural dynamics provides direct insights into transient lattice modifications induced by photoexcitation, shedding light on the carrier-lattice coupling during energy relaxation toward the band-edge states in 2D perovskite materials.

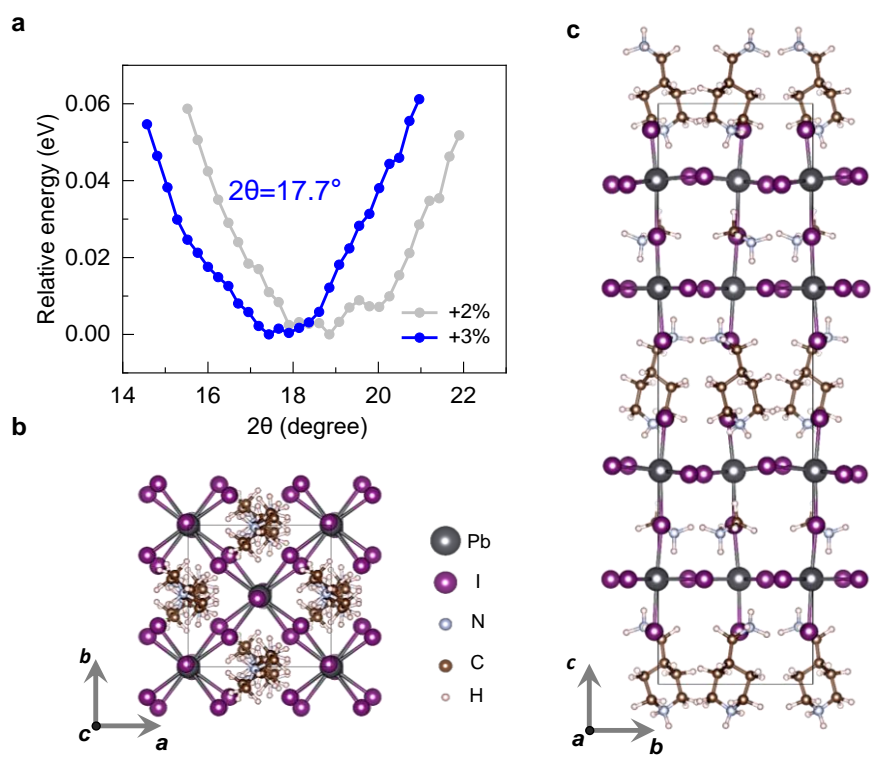

**Figure 4. Photostriction reduced structure distortion in 2D perovskite under photoexcitation**. (a) Energy profile of the 2D perovskite (4AMP)MAPb$_2$I$_7$ subjected to photodoping considering lattice expansions of both 2% and 3%. The energy minima for the 2% and 3% expanded lattices are observed at 18.8° and 17.7°, respectively. The laser intensity is E$_0$=0.75 V/Å. (b, c) Top and side views of snapshot for photodoping induced atomic structures, respectively.

The photostriction effect plays a crucial role in mitigating structural distortions in 2D perovskites under photoexcitation. Laser-induced lattice expansion can further modulate the distortion angles in 2D perovskite materials.

As shown in Figure 4a, the energy landscape exhibits a minimum at 2θ=18.9° for the 2% expanded lattice subjected to a laser field of $E_0$=0.75 V/Å and a photodoping-induced lattice expansion (grey line). For larger lattice with 3% expansion, the distortion reduces to 2θ=17.7° with the same laser pulses (blue curve). The structural snapshots from both top and side views reveal the dynamic atomic rearrangements driven by the photostriction effect (Figure 4b-c). Strong Laser pulses induce lattice expansion and effectively enhance the relaxation process of structural distortion, reduced on the scale of 26%. This suggests that laser excitation is essential for modulating lattice dynamics, potentially influencing structural stability and performance.

Therefore, the photostriction effect plays a crucial role in the relaxation of lattice distortion in 2D perovskites. The expanded lattice facilitates strain relaxation by increasing interatomic distances and alleviating internal stresses, ultimately stabilizing the structure. These studies highlight the potential of photostriction in tailoring strain engineering for the optoelectronic applications. In contrast, it is distinct from the Debye-Waller effect [2], which results in structural deviations from the equilibrium positions and leads to increased destructive interference and a loss of coherent scattering intensity. The emergence of photodoping carriers facilitates the relaxation of lattice distortions on femtosecond timescales, offering valuable guidance for the optoelectronic and energy-harvesting applications by revealing key structural mechanisms. Our findings provide critical insights into intrinsic photoinduced lattice dynamics, shedding light on the mechanisms underlying laser-driven structural modifications in photovoltaic materials.

In conclusion, we have demonstrated that the photostriction effect can facilitate relaxation of lattice distortion in 2D perovskites. Strong laser pulses introduce high densities of photodoping carriers in 2D perovskites and modulate the lattice stability. Using first-principles calculations, we reveal the photoexcited lattice expansion introduces structural relaxation (~20%) in 2D perovskites facilitated by the photostriction effect. This is attributed to strain relaxation by increasing interatomic distances and alleviating internal stresses, ultimately stabilizing the structure. Furthermore, the interaction between optically-excited carriers and lattice dynamics can be systematically tailored by adjusting the photodoping density, revealing a novel approach to tuning light-matter interactions in 2D perovskites. Thus, the study highlights the fundamental role of photostriction in engineering structural dynamics, paving the way for rational design strategies to optimize material performance in perovskite-based solar cells and optoelectronic devices.

**Acknowledgments**

This work was supported by the starting funding from National Center for Nanoscience and Technology. This work was supported by the National Key R&D Program of China (2022YFA1203200), the Basic Science Center Project of the National Natural Science Foundation of China (22388101), the Strategic Priority Research Program of the Chinese Academy of Sciences (XDB36000000), the National Natural Science Foundation of China (12125202). The numerical calculations in this study were partially carried out on the ORISE Supercomputer. J.Z. thanks Z. Ding for fruitful discussions.

**Author contributions**

J.Z. designed the research. All authors contributed to the analysis and discussion of the data and the writing of the manuscript.
**Conflict of Interest:** The authors declare no competing financial interest.

# Methods

   **Time-dependent density functional theory.** The TDDFT calculations were performed utilizing the time-dependent *ab initio* package (TDAP), developed based on the time-dependent density functional theory [35-37] and implemented within SIESTA [38-41]. The dynamic simulations were carried out with an evolving time step of 50 as for both electrons and ions within a micro-canonical ensemble. To explain the TDDFT methods more, we present more description on our methods based on time-dependent Kohn-Sham equations for coupled electron-ion motion. We perform ab initio molecular dynamics for coupled electron-ion systems with the motion of ions following the Newtonian dynamics while electrons follow the time-dependent dynamics. The ionic velocities and positions are calculated with Verlet algorithm at each time step. When the initial conditions are chosen, the electronic subsystem may populate any state, ground or excited, and is coupled nonadiabatically with the motion of ions. We carried out k-space integration using a 2×2×1 mesh for the bilayer case in the Brillouin zone of the supercell to confirm the convergence in real-time propagation.

**Constrained density functional theory calculations.** This method modifies the electron density or potential to enforce specific electronic configurations, enabling the exploration of excited states, charge transfer, and other phenomena not naturally favored by standard DFT. It provides a systematic way to explore electronic structures that are not naturally favored by standard DFT but are relevant for understanding key physical and chemical properties in constrained environments [42].

The electron density [$\rho^{OCDFT}(r)$] in the general form of an orbital-constrained DFT expression is:

$$\rho^{OCDFT}(r) = \sum_i n_i^{OCDFT} \phi_i^*(r)\phi_i(r) \quad (2)$$

The total energy for the orbital-constrained DFT is written as $E_{KS}[\rho^{OCDFT}(r)]$ instead of ground-state $E_{KS}[\rho_0(r)]$. Herein $E_{KS}[\rho]$ is based on the orbital-constrained electron density, and $\rho_0(r)$ is the electron density in the ground state $\rho_0(r) = \sum_i n_i \phi_i^*(r)\phi_i(r)$; $n_i$ is the occupation number of the *i*-th orbital in ground state and $n_i^{OCDFT}$ is the fixed target occupation for the *i*-th orbital for the excited states.